\documentclass[journal,comsoc]{IEEEtran}
\usepackage[T1]{fontenc}

\usepackage{cite}

\ifCLASSINFOpdf
   \usepackage[pdftex]{graphicx}
\else
\fi
\usepackage{amsmath}
\usepackage[cmintegrals]{newtxmath}
\usepackage{algorithmic}

\usepackage{amsfonts}

\usepackage{caption}
\usepackage{subcaption}
\usepackage{xcolor}

\begin{document}
%
\title{Wireless Image Retrieval at the Edge}
%
%
%

\author{Mikolaj Jankowski, Deniz G{\"u}nd{\"u}z and Krystian Mikolajczyk\\
Imperial College London
\thanks{This paper was presented in part at the 45th IEEE International Conference on Acoustics, Speech, and Signal Processing (ICASSP) \cite{jankowski2019deep}.}
\thanks{This work was supported in part by the European Research Council (ERC) through Starting Grant BEACON under Grant 677854 and in part by the U.K. Engineering and Physical Sciences Research Council (EPSRC) under Grant EP/N007743/1, Grant EP/S032398/1, and Grant EP/T023600/1. }
}

\maketitle

\begin{abstract}
We study the image retrieval problem at the wireless edge, where an edge device captures an image, which is then used to retrieve similar images from an edge server. These can be images of the same person or a vehicle taken from other cameras at different times and locations. Our goal is to maximize the accuracy of the retrieval task under power and bandwidth constraints over the wireless link. Due to the stringent delay constraint of the underlying application, sending the whole image at a sufficient quality is not possible. We propose two alternative schemes based on digital and analog communications, respectively. In the digital approach, we first propose a deep neural network (DNN) aided retrieval-oriented image compression scheme, whose output bit sequence is transmitted over the channel using conventional channel codes. In the analog joint source and channel coding (JSCC) approach, the feature vectors are directly mapped into channel symbols. We evaluate both schemes on image based re-identification (re-ID) tasks under different channel conditions, including both static and fading channels. We show that the JSCC scheme significantly increases the end-to-end accuracy, speeds up the encoding process, and provides graceful degradation with channel conditions. The proposed architecture is evaluated through extensive simulations on different datasets and channel conditions, as well as through ablation studies. 
\end{abstract}

%
\IEEEpeerreviewmaketitle

\section{Introduction}
%
%
%
%
\label{sec:introduction}

\IEEEPARstart{I}{nternet} of Things (IoT) devices are becoming increasingly widespread. These small specialized computers are present in  offices, streets, and homes. Their main goal is to continuously sense their environment, and send the measurements through a wireless channel to an edge server, which performs data collection and further processing. Typical approach in most IoT applications is to convey all the measurements from the  IoT devices to an edge server, where state-of-the-art machine learning algorithms are used to analyse the collected data. However, in some applications, the volume of the measurement data (e.g., images, videos or LIDAR data) is large, and transmitting it to the server at the required quality may not be feasible within the limited latency requirements, e.g., in autonomous driving, surveillance, drones, etc. On the other hand, as the computational capabilities of IoT devices advance, they can process the data locally before offloading it to a server. In some cases  the desired inference tasks can be carried out locally, which is beneficial as the IoT devices have access to the original data, rather than its quantized version at the edge server, due to the lossy compression and transmission over the wireless channels.

\begin{figure}[]
\begin{center}
\includegraphics[width=0.95\linewidth]{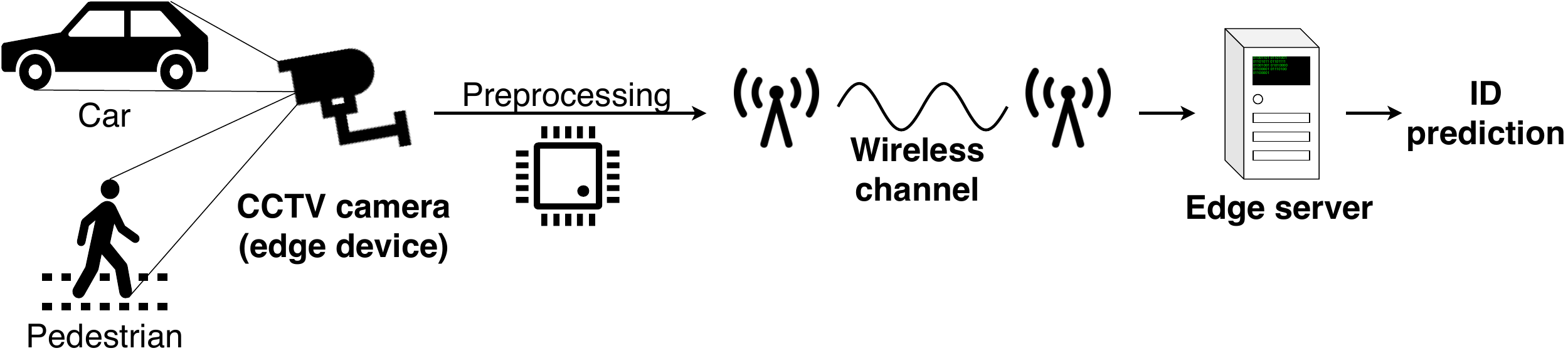}
\end{center}
  \caption{An illustration of the retrieval problem at the edge. A CCTV camera takes a picture of a pedestrian or a car, and processes the image locally to obtain a low-dimensional signature, which is then sent through a wireless channel to an edge server that performs identification based on a large database it has access to.}
\label{fig:system_diagram}
\end{figure}

In this work, we study machine learning at the wireless edge. In particular, we focus on distributed inference over a wireless channel, where a centrally-trained algorithm is deployed on IoT devices to perform inference over-the-air. One of the machine learning tasks for which remote inference is essential is retrieval. In autonomous vehicles, drones, or in surveillance systems, agents try to identify objects, vehicles, or humans in their environment through their sensory data. The goal in image retrieval is to identify a query image of a person or a vehicle recorded locally by matching with images stored in a large database (gallery), typically available at the edge server (cf. Fig.~\ref{fig:system_diagram}). We emphasize that the retrieval task cannot be performed locally at the edge device regardless of its computational power. This is because the centralized database is available only at the edge server, hence, some sensory data has to be transmitted to the edge server. The fundamental question we want to answer in this paper is what part or function of data must be transmitted, and how.

A trivial response to these questions would be to convey the image to the edge server at the best quality possible. The server first reconstructs the image, and performs the retrieval task with a state-of-the-art retrieval algorithm. Note, however, that a significant part of the image content may not be relevant for the retrieval task, therefore the original image is not needed at the server. Indeed, novel approaches to retrieval employ deep neural networks (DNNs) as feature encoders that map input images to a low-dimensional feature space, such that vectors extracted from the same identities are similar, despite different views or occlusions. Accordingly, we employ DNNs for extracting features that are then transmitted over the wireless link.

We propose two approaches to convey the feature vectors to the edge server. In the conventional ``digital'' approach, feature vectors are first compressed, and encoded with a channel code for reliable transmission. The features that are most relevant for the retrieval task  are extracted and transmitted depending on the capacity and the reliability of the channel between the edge device and the server. To improve the efficiency of this approach, we design a retrieval-oriented image compression scheme, which compresses the feature vectors depending on the available bit budget. This ``separate'' data compression and channel transmission scheme assumes reliable communication over the channel. Such scenario is typically difficult to achieve in practice, especially for short blocklengths considered in this work, imposed by the strict delay limitations. Alternatively, we consider a joint source and channel coding (JSCC) approach, where the feature vectors are directly mapped into channel input symbols, and the noisy channel output is used by the server to retrieve the most relevant images, without involving any explicit channel code. This can be considered as ``analog'' communication since the feature vectors are not converted into bits at any stage. For the JSCC approach, we employ an architecture based on DNNs, similar to the novel DeepJSCC \cite{jscc_image_dnn, jscc_feedback}, which has recently been  introduced for wireless image transmission. Our results show that the JSCC scheme can outperform the highly optimized feature compression scheme even if we assume the availability of capacity-achieving channel codes for the digital scheme. To the best of our knowledge, this is the first work to study image retrieval over a wireless channel. Our specific technical contributions can be summarized as follows:

\begin{itemize}
    \item We propose a novel retrieval-oriented image compression scheme, which combines a retrieval baseline with a feature encoder, followed by scalar quantization and entropy coding. To estimate the distribution to be used for the entropy coder, we introduce a density model based on a Gaussian mixture.
    \item We propose an autoencoder-based architecture and training strategy for robust JSCC of feature vectors, generated by a retrieval baseline, under noisy, fading, and bandwidth-limited channel conditions.
    \item We perform extensive evaluations under different signal-to-noise ratio (SNR) and bandwidth constraints, and show that the JSCC scheme outperforms the digital approach even with capacity-achieving channel codes. Moreover, its performance exhibits graceful degradation when the test and training SNRs do not match. The JSCC scheme is shown to outperform its digital counterpart also over fading channels, even if we assume the availability of channel state information for the digital scheme only. 
    \item We evaluate the proposed schemes on various surveillance tasks, and show that the performance close to the noiseless bound can be achieved even under very harsh SNR and bandwidth constraints, whereas the digital approach falls short of this performance even with idealistic capacity-achieving channel codes. 
    \item Our results show that, in general, it is  not possible to separate inference tasks from the communication scheme, and the end-to-end performance can be improved significantly by designing the communication and learning algorithms jointly. We provide a comprehensive analysis of different architectures and training strategies that will serve as solid baselines for future research in wireless edge learning.
\end{itemize}

In this paper we extend our previous work \cite{jankowski2019deep} by  considering different wireless channel models to show the generalization of our method and we validate the methods by extensive evaluations on new datasets. We provide a comparison of different architectures and training methods for wireless image retrieval. In our digital model we introduce a new, simpler, but equally effective density model based on a Gaussian mixture. 


\section{Related work}

\subsection{Machine Learning at the Wireless Edge}

With the increasing computational capabilities of edge devices, many recent studies consider executing machine learning tasks across edge devices. Many of these works focus on the training stage, which is particularly challenging due to the distributed nature of data available at edge nodes, and the typically limited communication resources (please see \cite{Quek:FL_schedule, Gunduz_JSAC, Park:Proc:19, amiri_wireless_edge, Amiri:TWC:20, Mehdi:ICASSP:20} and references therein). 

Instead, in this work, we focus on the inference phase, assuming that the training can be run centrally. This approach requires centralized availability of the training data. Prior works on distributed inference at the wireless edge have focused on classification tasks using DNNs. Authors of \cite{2step_pruning, bottlenet, bn_plus_plus, dnn_decoupling, jscc_features, jankowski_spawc} suggest splitting neural network architectures into two parts to reduce the computational workload at the edge device. In this work we do not consider computational limitations of the device, and perform the forward pass over the DNN locally, at the edge device, which was shown in \cite{jankowski_spawc} to reduce the bandwidth necessary to transmit the information for the classification task. Digital schemes for distributed inference, e.g. \cite{bottlenet, dnn_decoupling}, limit the amount of information (e.g., the number of bits) that can be conveyed to the edge server, but ignore the energy and latency cost of communications, and potential errors that may be introduced. However, in practice, reliable transmission of the feature vectors, even if they are highly compressed, requires an accurate estimate of the channel state at the edge device, and a very reliable error correction code. However, not only such a separate approach is suboptimal, but also channel codes introduce significant error probability at short blocklengths, especially in the absence of accurate channel state information. Analog schemes based on JSCC have recently been  considered in \cite{bn_plus_plus, jscc_features, jankowski_spawc}, and they were shown to outperform separate approaches, but they focus on the classification task using low-resolution images. This significantly reduces the amount of information to be transmitted, as the task is to distinguish between a finite set of known classes. In contrast, in the retrieval and re-identification tasks, we require high resolution images, and have to cope with unknown set of identities, thus the feature vectors have to convey significantly more information. Unlike in the classification, the retrieval task cannot be performed locally at the edge device due to its limited computational resources and data transmission to the edge server is needed.

\subsection{Person and Vehicle Retrieval}

Person and vehicle retrieval tasks have been extensively studied \cite{DBLP:journals/corr/ZhengYH16, he2019part,pose_invariant_reid, kuma2019vehicle,attention1,pyramid,zhou2018aware}. They share the same motivation to allow for a better and more reliable recognition of people and vehicles, mainly targeting surveillance applications. The most successful recent approaches for image retrieval problems are based on convolutional neural networks, and recent techniques include part classifiers \cite{DBLP:journals/corr/ZhengYH16, he2019part}, creating bias-invariant feature vectors \cite{pose_invariant_reid, zhou2018aware}, using attention models \cite{attention1}, and analyzing images at different scales \cite{mgn, multi_scale1}. Despite the popularity of triplet loss in both areas \cite{triplet, kuma2019vehicle}, designs based on softmax cross-entropy have also been successfully implemented \cite{pyramid}.

\subsection{Joint Source-Channel Coding (JSCC)}

According to Shannon's separation theorem \cite{shannon}, performing source and channel coding separately achieves theoretical optimality guarantees in the asymptotic infinite blocklength regime. This theorem holds under average power constraint and a single-letter additive distortion constraint, e.g., average mean-square error between the samples of the input and output sequences. However, in practice, we are limited by finite blocklengths due to complexity and latency constraints; and JSCC is known to outperform separate schemes in practical scenarios. Many JSCC schemes have been proposed \cite{jscc_video, jscc_images, jscc_wireless}, but these have not found application in practice as they are too complex and specific to the underlying source and channel distributions. Moreover, they do not provide sufficient improvement to justify the introduced increase in the system complexity, as well as the loss of modularity. More recently, JSCC schemes based on autoencoders \cite{autoencoders}, which are DNNs aimed at unsupervised data coding, have been introduced \cite{jscc_text, jscc_image_dnn, david_spawc, jscc_feedback}, and are shown to provide comparable or better performance than state-of-the-art digital schemes.

JSCC for remote inference problems is much less studied. Distributed hypothesis testing problem over a noisy communication channel has been recently introduced in \cite{Sreekumar:IT:20} using an information theoretic formulation and considering the type II error exponent as the performance measure. Here, the goal is to make a decision on the joint distribution of the samples observed by a remote observer and those observed by the decision maker. Similarly to our setting, the observer communicates to the decision maker over a noisy channel. It is shown that, while the optimality of separation holds for the problem of testing against independence, where the alternative hypothesis is the product of the marginal distributions of the remote and local samples, separation is suboptimal in general, when testing against arbitrary joint distributions. 


\section{Methods}

In this work we propose two approaches for performing retrieval over wireless channels: digital (separate) and JSCC (joint) approaches. In both cases, we consider the transmission of the feature vectors, which are a low-dimensional representation of identities of the items to be retrieved e.g., humans, vehicles (Section \ref{subs:prid_baseline}), and have to be sent over bandwidth-limited wireless channels. Due to the channel limitations, features cannot be transmitted in a lossless fashion, and have to be compressed. The recovered noisy feature vectors at the receiver are compared to the feature vectors of images previously collected from other edge cameras, called the \textit{gallery}, in order to find the nearest neighbour.

\subsection{Channel Model}
\label{subs:channel_model}
We assume that the edge device is connected to the edge server through an additive white Gaussian noise (AWGN) channel. We consider static as well as slow fading channel. For both approaches presented in this work, we assume that the channel model is known during training, and remains the same during inference.

The AWGN channel is characterized as follows: given a channel input vector $\mathbf{x} \in \mathbb{C}^B$, consisting of $B$ complex channel input symbols $x_i$, the output $\mathbf{y} \in \mathbb{C}^B$ is given by $\mathbf{y} = \mathbf{x} + \mathbf{z}$, where $z_i \sim \mathbb{C}\mathcal{N}(0, \sigma^2)$ are the independent and identically distributed (i.i.d.) elements of the noise vector $\mathbf{z} \in \mathbb{C}^B, i=1,\dots, B$. An average power constraint is imposed on the input vectors, such that $\frac{1}{B} \sum_{i=1}^B |x_i|^2 \leq P = 1$; which, in the case of a static AWGN channel, translates into a maximum received SNR of $\mathrm{SNR} = 10 \log_{10}(\frac{1}{\sigma^2})$ in $\mathrm{dB}$ scale.

In the slow fading scenario, we consider a single-tap Rayleigh fading channel model, where all the transmitted symbols experience the same channel gain. That is, given the channel input vector $\mathbf{x} \in \mathbb{C}^B$, the corresponding output vector $\mathbf{y} \in \mathbb{C}^B$ is given by $\mathbf{y} = h\mathbf{x} + \mathbf{z}$, where $h\sim \mathbb{C}\mathcal{N}(0, \sigma_h^2)$ and $z_i\sim \mathbb{C}\mathcal{N}(0, \sigma^2)$ are drawn from independent zero-mean complex normal distributions with variances $\sigma_h^2$ and $\sigma^2$, respectively. We impose the same average input power constraint of $P=1$ as in the AWGN case. For each transmitted feature vector we use a single gain $h$, which characterizes the \textit{slow fading} behaviour. The maximum average SNR is evaluated by $\mathrm{SNR} = 10 \log_{10}(\frac{\sigma_h^2}{\sigma^2})~\mathrm{dB}$, while for all the experiments shown in this paper we set $\sigma_h^2=1$, which corresponds to the same average received power as in the static AWGN channel model.

\subsection{Retrieval Baseline}
\label{subs:prid_baseline}

Following the state-of-the-art retrieval methods \cite{pyramid, he2019part} we employ the ResNet-50 network \cite{resnet}, pretrained on ImageNet \cite{imagenet}, for feature extraction. This ensures that similar results can be expected in different setups. In more detail, we use ResNet-50 with batch normalization (BN) layers applied after each convolutional layer. As input, we use images resized to a common $256 \times 128$ resolution with bicubic interpolation for person datasets and $128 \times 128$ resolution for vehicle datasets. For the last layer we use average pooling across all the feature maps, which results in a 2048-dimensional feature vector. During training we use stochastic gradient descent (SGD) with a learning rate of $0.01$ and a momentum of $0.9$. We also apply $L_2$ regularization, weighted by $5\cdot 10^{-4}$ to the ResNet-50 parameters. We refer to this architecture as the \textit{feature encoder}.

\subsection{Digital Transmission of Compressed Feature Vectors}
\label{subs:approach3}

\begin{figure*}[]
\begin{center}
\includegraphics[width=\linewidth]{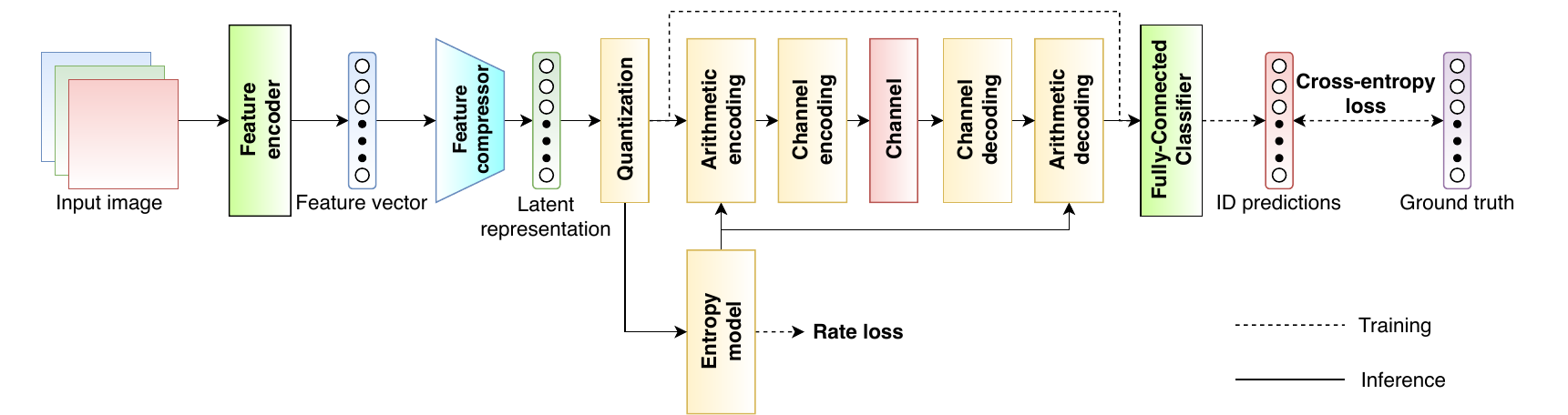}
\end{center}
  \caption{The digital transmission scheme. Input is transformed into a feature vector, which is compressed using a DNN. At the receiver, latent representation is classified into IDs to compute the loss during training only. Arithmetic coding and channel coding is bypassed during training.}
\label{fig:approach3}
\end{figure*}

This approach is based on the assumption that a certain number of bits can be reliably conveyed to the edge server for each image. In practice, however, this is highly challenging to achieve. Ultra-reliable channel codes require large blocklengths even in the static AWGN setting, and accurate channel estimation and feedback in the slow-fading case. In our simulations, we assume capacity-achieving channel codes, which will serve as a bound on the performance of practical digital schemes.

An overview of the proposed digital scheme is shown in Fig.~\ref{fig:approach3}.  We first extract features using the retrieval baseline described in Section \ref{subs:prid_baseline} as feature encoder. The resulting feature vector is compressed into as few bits as possible through lossy compression followed by arithmetic coding. The compressed bits are then channel coded, with introduced structured redundancy to combat channel impairments. 

The lossy feature compressor consists of a single fully-connected layer for dimensionality reduction, followed by quantization. On the receiver side we use the quantized latent representation as a feature vector, which is passed through a fully-connected layer for ID classification. Note that the IDs and their classification are used for calculating the loss during training only. During retrieval, the IDs are not known and the feature vectors are used for nearest neighbour search. This has been shown to perform well in the re-ID community \cite{DBLP:journals/corr/ZhengYH16, he2019part,pose_invariant_reid, kuma2019vehicle,attention1,pyramid,zhou2018aware}.

To enable an end-to-end differentiable approach, we utilize the well-known quantization noise \cite{quantization} to model the quantization process. Specifically, instead of rounding the latent representation to the nearest integer, in the training phase we add the uniform noise to each element of the latent representation as follows:
\begin{equation}
    Q(\mathbf{z}) = \mathbf{z} + \mathcal{U}\left(-\frac{1}{2}, \frac{1}{2}\right),
\end{equation}
where $Q(\cdot)$ is the approximated quantization operation, $\mathbf{z}$ is the latent representation, and $\mathcal{U}(\cdot, \cdot)$ is the uniform noise vector. This formulation ensures a good approximation of quantization during training, whereas we perform rounding to the nearest integer during inference.

In order to optimize the arithmetic coder, we estimate the distribution of the quantized outputs. We assume that the elements $q$ of vector $\mathbf{q}=Q(\mathbf{z})$ are i.i.d. with some probability mass function (PMF) $p(q)$. To model this PMF, we propose a simple yet flexible solution using a mixture of Gaussians. We first approximate $p(q)$ as a continuous-valued probability density function $p_c(q)$ as follows:

\begin{equation}
    p_c(q) = \sum_{k=1}^K \alpha_k \frac{1}{\sigma_k \sqrt{2\pi{}}} e^{-\frac{1}{2}\left(\frac{q-\mu_k}{\sigma_k}\right)^2},
\end{equation}
where $K$ is the number of mixtures, $\sigma_k$ are mixture scales, $\mu_k$ are mean values, and $\alpha_k$ are the corresponding mixture weights. In our experiments we set $K=9$, which we empirically found to perform the best. Then, in order to evaluate our PMF $p(q)$ at discrete values $q \in \mathbb{Z}$, we integrate $p_c(q)$ over $\left[q-\frac{1}{2}, q+\frac{1}{2}\right]$ to obtain:
\begin{equation}
p(q) = \int_{q - \frac{1}{2}}^{q + \frac{1}{2}}p_c(x) dx = 
F_c\left(q + \frac{1}{2}\right) - F_c\left(q - \frac{1}{2}\right),
\end{equation}
where $F_c$ is the cumulative density function of the distribution $p_c(q)$.

We remark that, here we learn the distribution of the quantized feature vectors, but unlike recent works \cite{balle_hierarchical, mentzer_lossless_cvpr} , we do not consider adaptive probability model and do not introduce another neural network to predict parameters $\{\alpha_k, \mu_k, \sigma_k\}$ of the mixture. The reason for that is the proposed simple model performs sufficiently well, and we want to avoid introducing any communication overhead by sending additional parameters per image.  Instead, we use the available bandwidth for sending  quantized feature vectors only.

With the model presented above, we can easily estimate the PMF of the quantized vector $\mathbf{q}$, which can be directly used to feed the arithmetic coding engine in the test phase, but also to evaluate the average approximate entropy over the dataset in our loss function, which we define as a weighted sum of two objectives:

\begin{equation}
    L = l_{ce} - \lambda \cdot \log_2p(\mathbf{q}),
\end{equation}
where $l_{ce}$ is the cross-entropy between the predicted class (identity) and the ground truth for the retrieval task. The second component of the loss function corresponds to the empirical Shannon entropy of the quantized vector, representing the average length of the output of the arithmetic encoder. Such formulation allows for a smooth transition between the retrieval accuracy and number of bits necessary to send the feature vector in a lossy fashion. Moreover, minimizing the entropy term is equivalent to maximizing the likelihood of $p(q)$, which corresponds to increasing the certainty of our model, and allows a satisfactory fit of our approximated distribution to the true underlying distribution of the discrete symbols.

\begin{figure*}[]
\begin{center}
\includegraphics[width=0.95\linewidth]{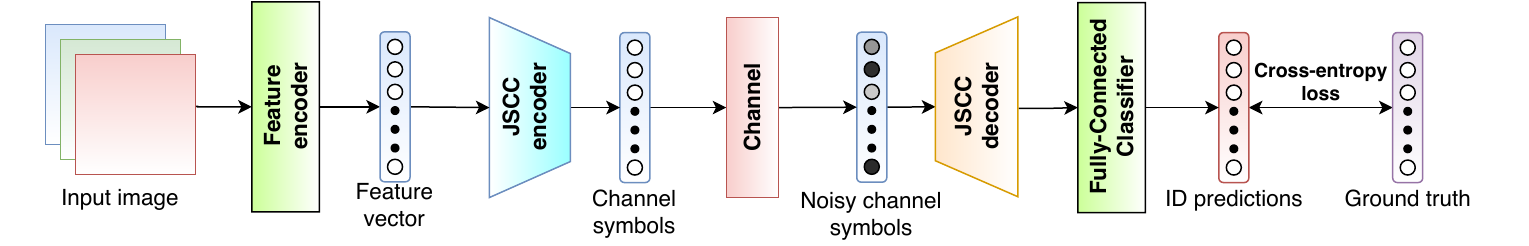}
\end{center}
   \caption{The architecture and training of JSCC of feature vectors for wireless image retrieval. The feature vector is directly mapped to channel inputs.  Received noisy signal is decoded and processed by a fully-connected layer to obtain ID predictions, which are then compared to the ground truth by the cross-entropy loss.}
\label{fig:approach1}
\end{figure*}

We apply the same settings discussed in Section~\ref{subs:prid_baseline} to train the feature encoder, the fully-connected classifier and the density model. 
We train the whole network for 20 epochs, reduce the learning rate to $0.001$ and train for further 30 epochs. 
We initialize our mixture parameters as follows: $\alpha_k = \frac{1}{K}$, $\mu_k = 0$, $\sigma_k = k^2, k=1, 2, \dots K$. To ensure the convergence during training, in the first epochs we prioritize the $l_{ce}$ loss term by setting the weight parameter $\lambda_i$ at epoch $i$ as:

\begin{equation}
    \lambda_i = \min \left(\lambda \frac{i}{E-20}, \lambda\right),\ i=1,\dots, E,
\end{equation}
where $E > 20$ is the total number of epochs. In our experiments we use $\lambda \in [10^{-5}, 10^{-2}]$, and $E=50$.

In the inference phase we use the arithmetic encoding engine to transmit the information with a channel code. Note that any channel code can introduce errors, there is therefore an inherent trade-off between the compression rate and the channel coding rate under a given constraint on the channel bandwidth, i.e., the number of channel symbols that can be transmitted to the edge server per image pixel. Compressing the feature vector further leads to increased distortion, and hence, reduced retrieval accuracy, but  also allows to introduce more redundancy, and hence, increased reliability against noise. In general, the optimal compression and channel coding rates depend on the retrieval accuracy-compression rate function of the compression scheme and the error-rate of the channel code. To simplify this task, we assume  capacity-achieving channel codes, which  provides an upper bound on the performance that can be achieved by any digital scheme that uses the above architecture. 

\subsection{JSCC of Feature Vectors}
\label{subs:approach1}

\begin{figure}[]
    \centering
    \begin{subfigure}[b]{0.37\linewidth}
        \centering
        \includegraphics[width=\textwidth]{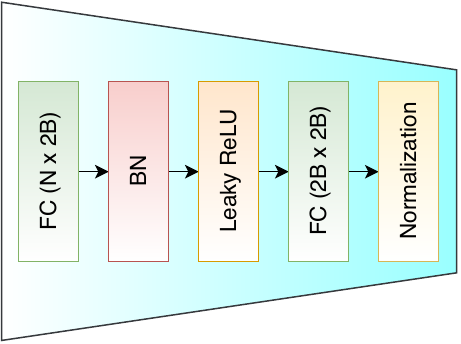}
        \caption{JSCC encoder}
        \label{fig:encoder}
    \end{subfigure}
    \hspace{10pt}
    \begin{subfigure}[b]{0.52\linewidth}
        \centering
        \includegraphics[width=\textwidth]{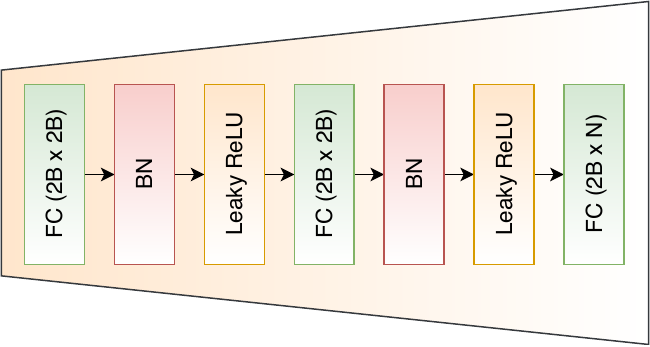}
        \caption{JSCC decoder}
        \label{fig:decoder}
    \end{subfigure}
   
    \caption{Proposed JSCC encoder and JSCC decoder architecture for the JSCC scheme illustrated in Fig.~\ref{fig:approach1}. At the encoder, dimensionality reduction is performed by the first fully-connected layer, which is inverted at the decoder.}
    \label{fig:autoencoder}
\end{figure}

In this section, we propose an alternative JSCC approach, called JSCC AE, and illustrated in  Fig.~\ref{fig:approach1}.  We use the baseline feature encoder as before to produce the feature vector for a given query image. The feature vector is mapped directly to the channel input symbols via a multi-layer fully-connected JSCC encoder (Fig.~\ref{fig:encoder}). We set the dimensionality of the channel input vector to $2B$ real symbols, which corresponds to the available channel bandwidth of $B$ complex values. In this work we consider small values of $B$ modeling stringent delay constraints of the underlying surveillance applications. This low-dimensional representation is normalized to satisfy the average power constraint of $P=1$, and transmitted over the AWGN channel. The noisy channel output vector at the receiver is mapped back to the high-dimensional feature space by a JSCC decoder (Fig.~\ref{fig:decoder}). The distance between the query feature vector and the feature vectors stored in the gallery set is calculated to find the nearest neighbours.

In order to train our network, the most straightforward strategy would be to perform end-to-end training, taking images from the dataset as an input, and training both the feature encoder and the JSCC autoencoder jointly, in an end-to-end fashion, with cross-entropy loss between the ID predictions and the ground truth (as shown in Fig. 3). However, our experimentation in Section IV-F shows that this approach leads to suboptimal performance. Alternatively, we propose traininig each component of the network separately at first, and, once the feature encoder and the JSCC autoencoder are pretrained individually, they are combined and trained jointly.
Therefore, our training strategy, which we refer to as $T_{1, 2, 3}$, consists of three steps: feature encoder pretraining ($T_1$), JSCC autoencoder pretraining ($T_2$), and end-to-end training ($T_3$). In the first step, $T_1$, we attach a single fully-connected layer at the end of the feature encoder that maps $2048$-dimensional feature vectors directly to the ID predictions. We pretrain the feature encoder for 30 epochs with a batch size of 16, using cross-entropy between the ID predictions and the ground truth as the loss function. In the second step, $T_2$, we freeze the pretrained feature encoder, and use it to extract features from all the images in the training dataset. We use these features as inputs to the proposed autoencoder network. We train the autoencoder using the $L_1$-loss between the feature vectors and the vectors reconstructed by the JSCC decoder. It is trained with SGD for 200 epochs with a learning rate $0.1$, reduced to $0.01$ after 150 epochs, and momentum of $0.9$. We apply $L_2$ regularizer to the autoencoder model, weighted by $5\cdot 10^{-4}$. Finally, in the third step, $T_3$, we train the whole network jointly, the autoencoder and the feature encoder, for 30 epochs, using the cross-entropy loss with a learning rate $0.001$, and for further 10 epochs with a learning rate of $0.0001$, applying the same optimizer and $L_2$ regularization as in the previous two steps.


Along with $T_{1, 2, 3}$ we evaluate four alternative training strategies. The first one, denoted by $T_{3}$, corresponds to the end-to-end training of the entire network (feature encoder + JSCC autoencoder + classifier) in a single training step. 
The second method, $T_{1,2}$, consists of the feature encoder pretraining, $T_1$, followed by the JSCC autoencoder training, $T_2$ to reconstruct feature vectors with $L_1$ as the distortion measure. This method corresponds to using a JSCC scheme whose goal is to reconstruct the feature vector as reliably as possible without taking into account the accuracy of the retrieval task. After $T_2$, the feature encoder and the autoencoder are combined as in Fig. 2, but the joint training step, $T_3$, is not performed.
The third method, $T_{1, 3}$, consists of the feature encoder pretraining, $T_1$, followed by joint training of the entire network, $T_3$. Finally, $T_{1, 3} + L_1$ approach is different from the $T_{1, 3}$  in that it combines the cross-entropy loss and $L_1$ loss, in the joint training phase.


Note that, we opted for an architecture that employs a distinct feature encoder and a separate JSCC autoencoder to transmit the feature vector over the channel. We have then trained these components in multiple training steps. It is possible to introduce a simpler architecture with a single JSCC encoder at the edge device that maps the query image to the channel input vector. Thus, no decoding is required at the receiver, and the retrieval task is directly performed using the noisy channel symbols. To compare our method to this straightforward approach, we introduce JSCC FC, which follows the same structure as in Fig.~\ref{fig:approach1}, except that the JSCC encoder is replaced by a single fully-connected layer and the  JSCC decoder is removed. We train the whole network end-to-end for $50$ epochs with cross-entropy loss, learning rate of $0.01$, reduced to $0.001$ after 30 epochs, and a momentum of $0.9$. We also apply $L_2$ regularization, weighted by $5\cdot 10^{-4}$, to all the parameters, including ResNet-50, feature encoder and fully-connected classifier.

\section{Results}

In this section we evaluate the performance of the proposed JSCC AE and JSCC FC architectures, and compare with that of the digital scheme presented in Section \ref{subs:approach3}, as well as the \textit{ideal channel} scenario with unlimited channel resources, where full, noiseless feature vectors can be transmitted over the channel. We first discuss the experimental setup and the dataset used for the evaluations.

\subsection{Experimental Setup}

\begin{figure*}[]
    \centering
    \begin{subfigure}[t]{0.49\textwidth}
        \centering
        \includegraphics[width=\textwidth]{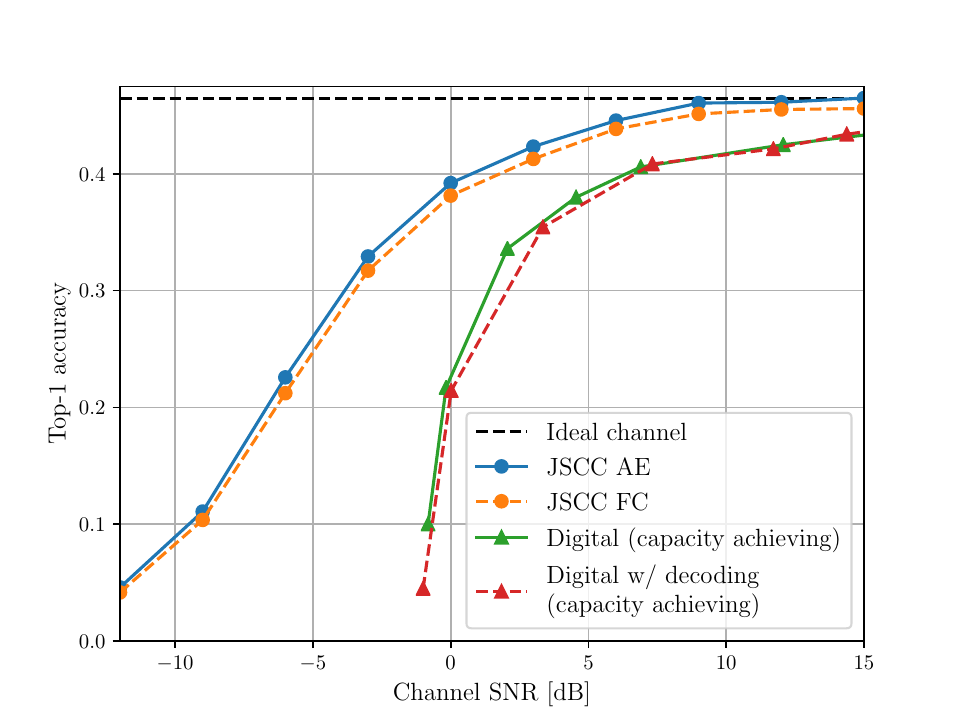}
        \caption{Person re-ID CUHK03 - AWGN}
        \label{fig:cuhk_awgn}
    \end{subfigure}
    \begin{subfigure}[t]{0.49\textwidth}
        \centering
        \includegraphics[width=\textwidth]{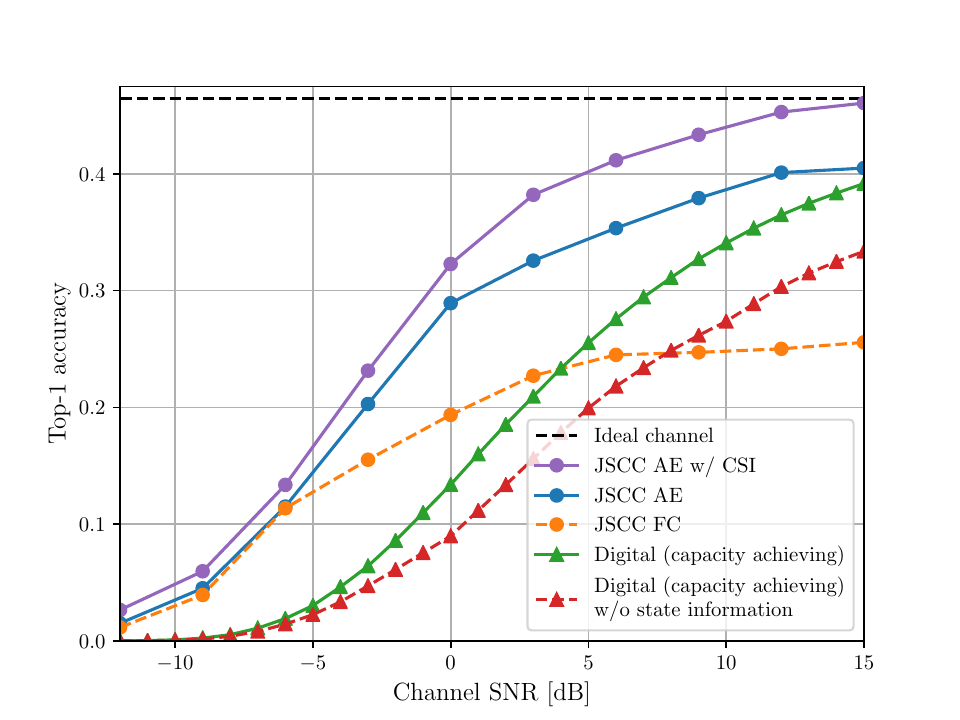}
        \caption{Person re-ID CUHK03 - fading}
        \label{fig:cuhk_fading}
    \end{subfigure}
    \begin{subfigure}[b]{0.49\textwidth}
        \centering
        \includegraphics[width=\textwidth]{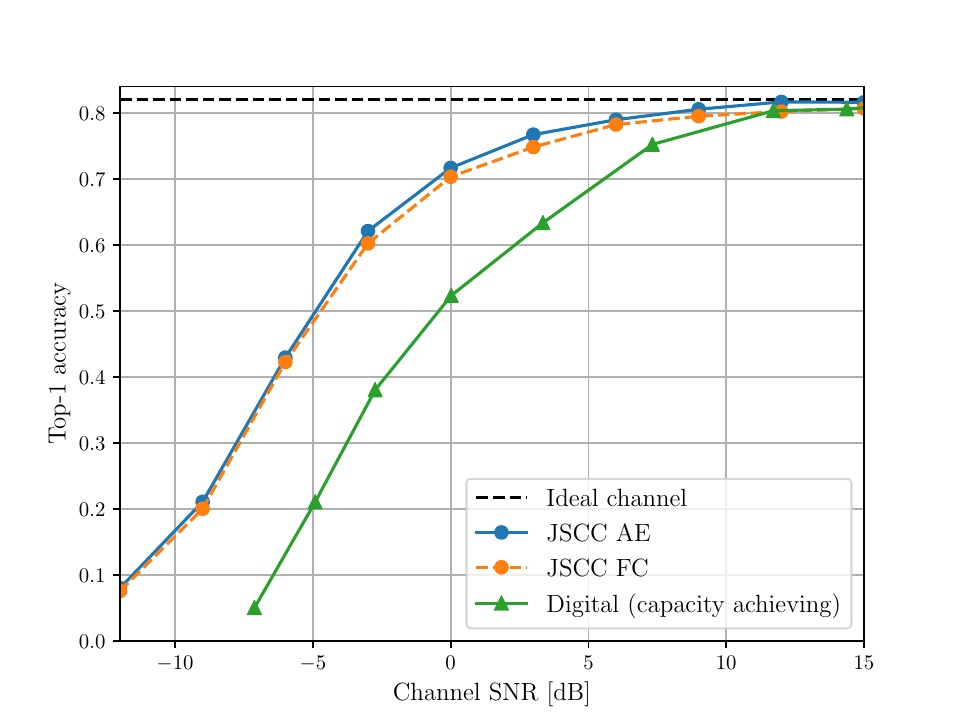}
        \caption{Person re-ID Market-1501 - AWGN}
        \label{fig:market_awgn}
    \end{subfigure}
    \begin{subfigure}[b]{0.49\textwidth}
        \centering
        \includegraphics[width=\textwidth]{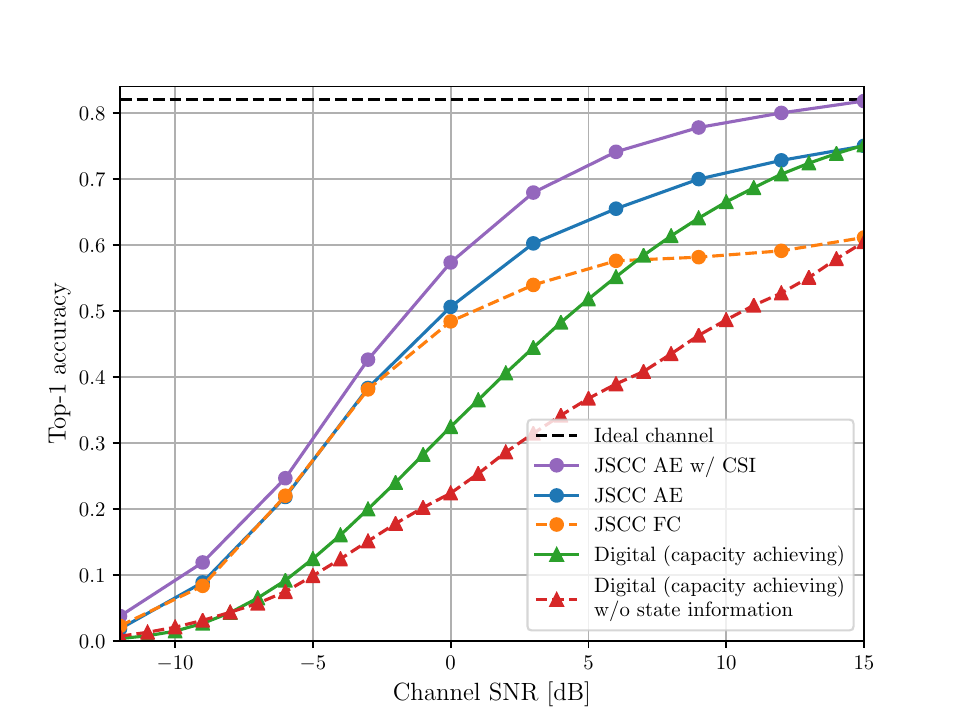}
        \caption{Person re-ID Market-1501 - fading}
        \label{fig:market_fading}
    \end{subfigure}
    \begin{subfigure}[b]{0.49\textwidth}
        \centering
        \includegraphics[width=\textwidth]{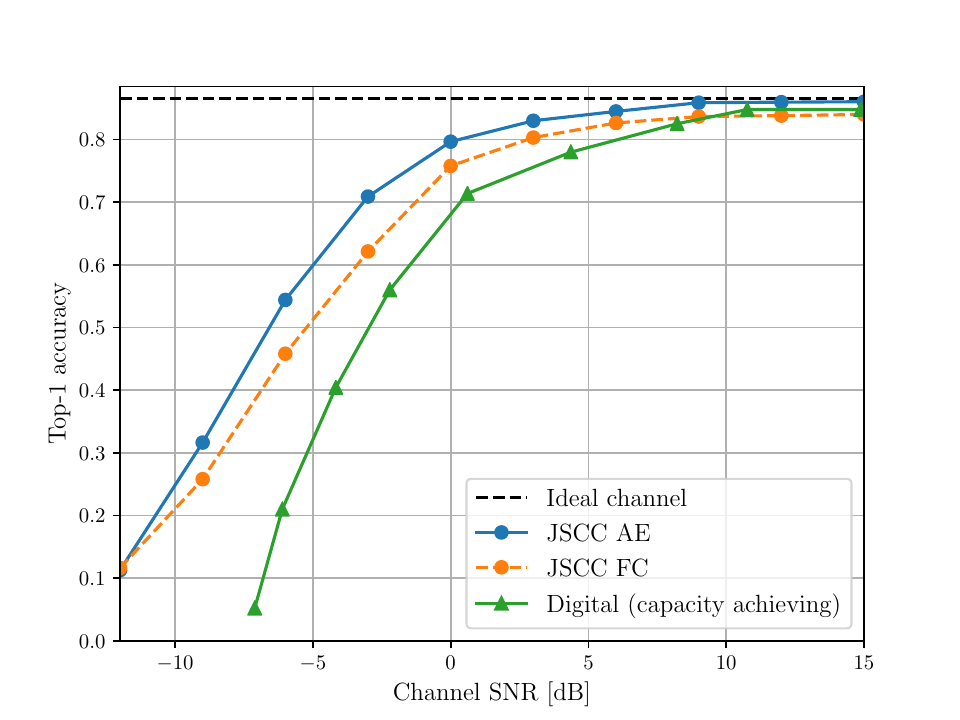}
        \caption{Car re-ID VeRi - AWGN}
        \label{fig:veri_awgn}
    \end{subfigure}
    \begin{subfigure}[b]{0.49\textwidth}
        \centering
        \includegraphics[width=\textwidth]{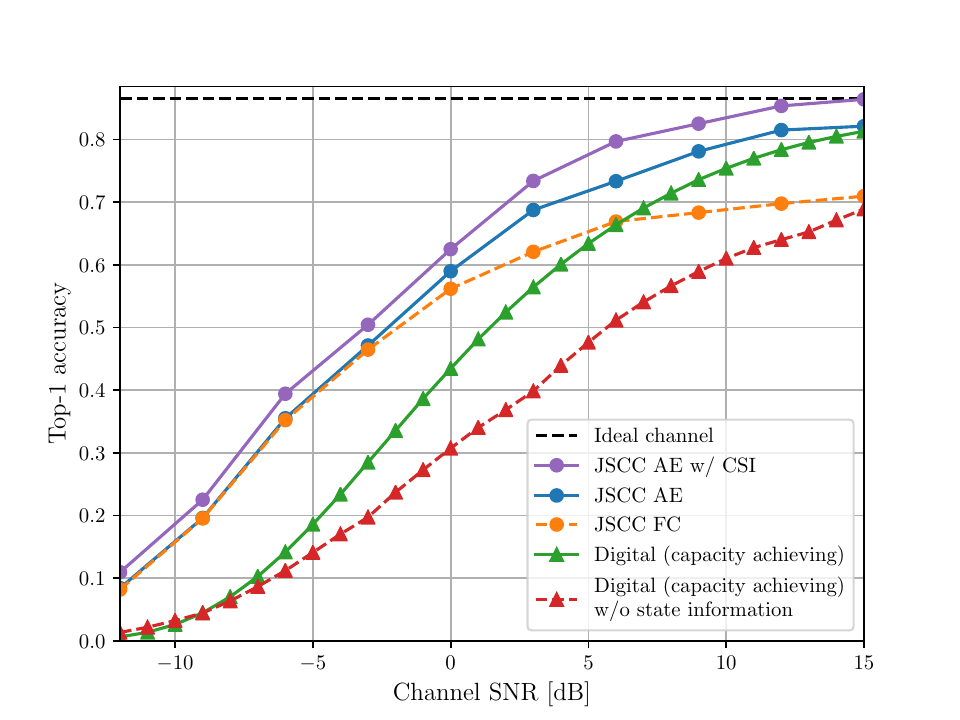}
        \caption{Car re-ID VeRi - fading}
        \label{fig:veri_fading}
    \end{subfigure}
    \caption{Performance comparison of the proposed three schemes over AWGN and slow fading channels for a range of channel SNRs and bandwidth  $B=64$. Our JSCC AE scheme achieves the best retrieval accuracy over the whole range of tested SNRs and for all three re-ID image retrieval datasets.}
    \label{fig:methods_comparison}
\end{figure*}

For the JSCC AE and JSCC FC schemes we vary channel SNRs for training,  between $\text{SNR}_{train} = -12\mathrm{dB}$ and $\text{SNR}_{train} = \infty \mathrm{dB}$, which corresponds to zero noise power. Training and test SNRs  are the same unless stated otherwise. In the digital scheme, we experiment with different dimensionality of the latent representation, between $64$ and $512$, estimate and minimize its entropy in the training phase by varying the value of parameter $\lambda$. In the testing phase we perform rounding to the nearest integer on each element of the latent representation and arithmetic coding, which is based on the probabilistic model learned by the entropy estimator, as described in Section \ref{subs:approach3}. This model assigns a probability estimate to each quantized symbol, which is then passed to the arithmetic encoder. We note that the proposed digital scheme is a variable-length encoder. Therefore, for a given fixed communication rate to the server, one has to determine the $\lambda$ coefficient that meets the rate constraint for each image. Instead, we fix the $\lambda$ coefficient and calculate the average number of bits required to encode the latent representations of the test images. We then evaluate the corresponding channel SNR to deliver these many bits to the receiver, assuming capacity-achieving channel codes. This is the upper bound on the real performance as practical codes are far from the capacity bound in the short blocklength regime. This model may correspond to sending multiple images together, and hence, the performance is determined by the average rate across many test images, rather than their individual rates.

For digital transmission over a fading channel, we consider two scenarios. In the first one, we assume perfect channel state information available at both the transmitter and the receiver. Then, for each query image and a corresponding random channel gain, we identify the $\lambda$ parameter that results in a bit rate that is as close as possible from below to the corresponding channel capacity. Then, we find the average accuracy across many random queries and channel conditions, following the underlying fading distribution. In the second scenario, we fix the $\lambda$ parameter, and for each query image and the corresponding random channel condition, we compare the required bit rate of the query image and the channel capacity. If the capacity is lower than the bit rate required by the compression scheme, we assume the transmission is failed. We then calculate the fraction of successful transmissions and multiply it by the average accuracy of the queries whose compressed feature vector can be successfully transmitted, for a given $\lambda$. Note that, there is a trade-off between the accuracy loss due to compression and the outage over the channel. 
The higher $\lambda$ values results in more compact representations of the feature vector, and hence less accurate retrieval performance even if they can be successfully conveyed to the server. Higher $\lambda$ values relaxes the compression constraint, but  may result in higher loss over the channel. Note that, we report only the results for the $\lambda$ values that lead to the highest average accuracy for each average SNR.

To train our model we used NVIDIA GeForce RTX 2080Ti GPU. A single end-to-end training of our digital model took approximately $35$ minutes, which was similar to the training time of the JSCC FC. For JSCC AE, the training took approximately 20 minutes, 3 minutes, and 30 minutes for $T_1$, $T_2$, and $T_3$, respectively. Please note that $T_1$ has to be performed only once, as this step does not depend on the channel model.

\subsection{Datasets}
In order to measure the performance of the retrieval task, we employ three widely used datasets:

\textbf{CUHK03} \cite{cuhk03} is a benchmark for person retrieval that contains $14096$ images of $1467$ identities taken from two different camera views. The  dataset was captured with six surveillance cameras and each identity within the dataset is represented by an average of 4.8 images per each of the two camera views. We use the \textit{labeled} variant of the dataset, where each image of the pedestrian was manually cropped by a human.

\textbf{Market-1501} \cite{market} contains $32217$ images of $1501$ pedestrians taken from a total of six cameras in front of a supermarket at Tsinghua University. Five out of six cameras are high-resolution cameras and the remaining one is low-resolution. Training and testing splits proposed by the authors contains $12936$  and $19732$ images, respectively. $750$ identities are additionally selected as a query set containing $3368$ images (maximum of $6$ per person). The dataset is different from CUHK$03$ in that it contains junk images capturing only partial pose and distractors presenting small fragments of pedestrian appearance or irrelevant objects.

\textbf{VeRi} \cite{veri1, veri2} is a vehicle retrieval dataset. It contains over 50000 images of 776 vehicles captured by 20 cameras within 24 hours over the area of  $1$km$^2$. Each identity is captured by 2-18 cameras in different viewpoints, occlusions, resolutions, and lighting conditions. All the images within the dataset are annotated with attributes, brands and colors, but in this work we do not utilize this information, and focus on retrieving the identity only based on the image.

The evaluation measure for all the datasets is the top-1 retrieval accuracy, which calculates the fraction of correct IDs at the top of the ranked list retrieved for each query.

\begin{figure*}[]
    \centering
    \begin{subfigure}[t]{0.49\textwidth}
        \centering
    \includegraphics[width=\textwidth]{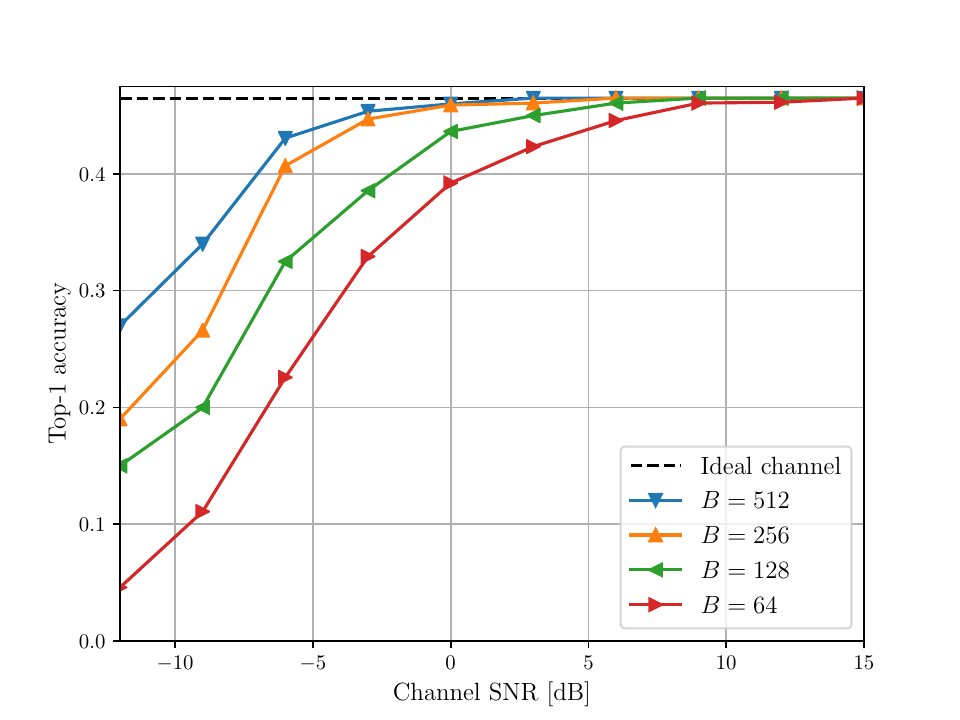}
    \caption{AWGN channel}
    \label{fig:results_bandwidth_cuhk_awgn}
    \end{subfigure}
    \begin{subfigure}[t]{0.49\textwidth}
        \centering
    \includegraphics[width=\textwidth]{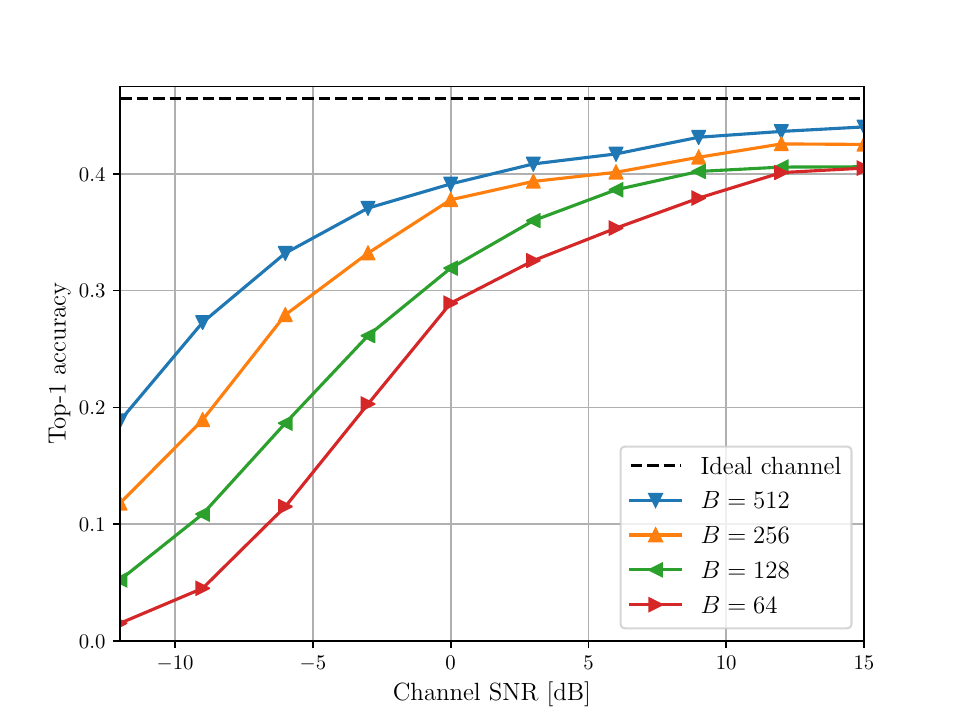}
    \caption{Fading channel}
    \label{fig:results_bandwidth_cuhk_fading}
    \end{subfigure}
    \caption{Accuracy  as a function of the channel SNR for different channel bandwidths. Higher bandwidth introduces more robustness against the channel noise.}
    \label{fig:results_bandwidth}
\end{figure*}

\subsection{Performance for Different Methods}
\label{subs:performance_comparison}

We plot the accuracy achieved by various schemes as a function of the test SNR in  Fig.~\ref{fig:methods_comparison}. For these experiments we use the bandwidth of 64, which corresponds to the transmission of 64 complex symbols through the channel. One can see that JSCC AE outperforms the digital scheme in all considered scenarios. For CUHK03 dataset the digital approach is not able to recover the noiseless accuracy even at $\mathrm{SNR}=15\mathrm{dB}$, whereas the proposed JSCC AE scheme obtains accuracy close to the ideal channel baseline at around $10\mathrm{dB}$ for the AWGN channel. JSCC FC follows JSCC AE very closely, but the increase in the performance provided by the autoencoder is visible for all the SNRs considered, which proves the superiority of the proposed architecture in comparison to the relatively simpler JSCC FC. The lower accuracy of JSCC FC may stem from the fact that the noise directly affects the low-dimensional feature vector, while the autoencoder-based scheme introduces certain level of denoising, which improves the feature estimates at the receiver. In Fig.~\ref{fig:cuhk_awgn}, we also show that feature decoding is not beneficial for the digital scenario. An alternative scheme, which we called Digital w/ decoding (capacity achieving), follows the same training strategy as discussed in Section \ref{subs:approach3}, but we further introduce a fully-connected decoder. This decoder is placed before the fully-connected classifier, and maps low-dimensional quantized latents back to the original, 2048-dimensional feature vector space. We show that this decoding step brings no improvement to the digital scheme performance, compared to the scenario without decoding. This result was consistent across all the datasets, but to avoid clutter we  show it only in Fig.~\ref{fig:cuhk_awgn}. Another observation is that the relative performances of the three schemes are similar for all the datasets considered, while JSCC FC seems to perform worse for the Car VeRi dataset, and even surpassed by the digital scheme at $\mathrm{SNR}=10\mathrm{dB}$.

Fading channels introduce additional perturbation to the channel symbols, reducing the accuracy of all the proposed approaches. Similarly to the AWGN channel, JSCC AE achieves the best performance across all three datasets and the average SNR values considered in this paper. The digital scheme performs worse when the channel state information is not available (which is also the case for the JSCC schemes). We have also included the performance of the digital scheme when perfect channel state information is available. We observe that even in this case the proposed JSCC AE scheme outperforms the digital alternative. 
JSCC FC closely follows JSCC AE at the low SNR regime, but its performance saturates to a level significantly below that of JSCC AE, and even  below the digital scheme for the CUHK03 dataset. This result further validates the denoising interpretation of the autoencoder structure in JSCC AE, which becomes even more critical in recovering the noisy feature vector in the presence of channel fading. Fading not only applies random attenuation to the received signal strength, but also random rotations in the complex plain, which makes it very difficult for the receiver to recover the features for correct retrieval without any channel state information. We note that, while the digital scheme suffers significant performance loss in the absence of the channel state information, JSCC AE seems to perform reasonably well. We can argue that the autoencoder learns to mitigate the effect of random fading despite the lack of explicit pilot signals. We also provide the performance of JSCC AE, when perfect channel state information is available at the receiver. The JSCC decoder first divides the received signal by the channel gain: $\mathbf{\hat{y}} = \frac{\mathbf{y}}{h} = \mathbf{x} + \frac{\mathbf{z}}{h}$, and reconstructs the input feature from $\mathbf{\hat{y}}$. In this scenario, our method is able to recover the noiseless bound at a reasonable $\mathrm{SNR}$ of $15\mathrm{dB}$, and the gap between JSCC AE w/ CSI and the digital approach grows even further.


\subsection{Performance for Different Bandwidths}

In this experiment we investigate the effect of the channel bandwidth $B$ on the retrieval performance for the person retrieval CUHK03 dataset, achieved by the JSCC AE scheme. We emphasize that the previously considered bandwidth of $B=64$ is extremely limited, corresponding to extremely low-latency communications, which may be essential for many surveillance and security applications. The top-1 accuracy as a function of the channel SNR is plotted in Fig.~\ref{fig:results_bandwidth} for different channel bandwidth values of $64, 128, 256$, and $512$. It shows that the accuracy and robustness increases significantly with the bandwidth, but the relative gain becomes smaller as we approach the original feature vector dimension.

For the fading channel,  it is visible in Fig.~\ref{fig:results_bandwidth_cuhk_fading} that the proposed JSCC AE scheme without the channel state information is not able to recover the original accuracy even for a significant bandwidth, and reaches a plateau at around $\mathrm{SNR} = 12 \mathrm{dB}$. As pointed out in Section \ref{subs:performance_comparison}, this may stem from the fact that our approach cannot fully cancel the effect of the variable channel gain. Channel estimation and feedback techniques can be utilized to mitigate the impact of random channel fading, as shown in Section \ref{subs:performance_comparison}.

\subsection{Graceful Degradation}

\begin{figure*}[]
    \centering
    \begin{subfigure}[t]{0.49\textwidth}
        \centering
    \includegraphics[width=\textwidth]{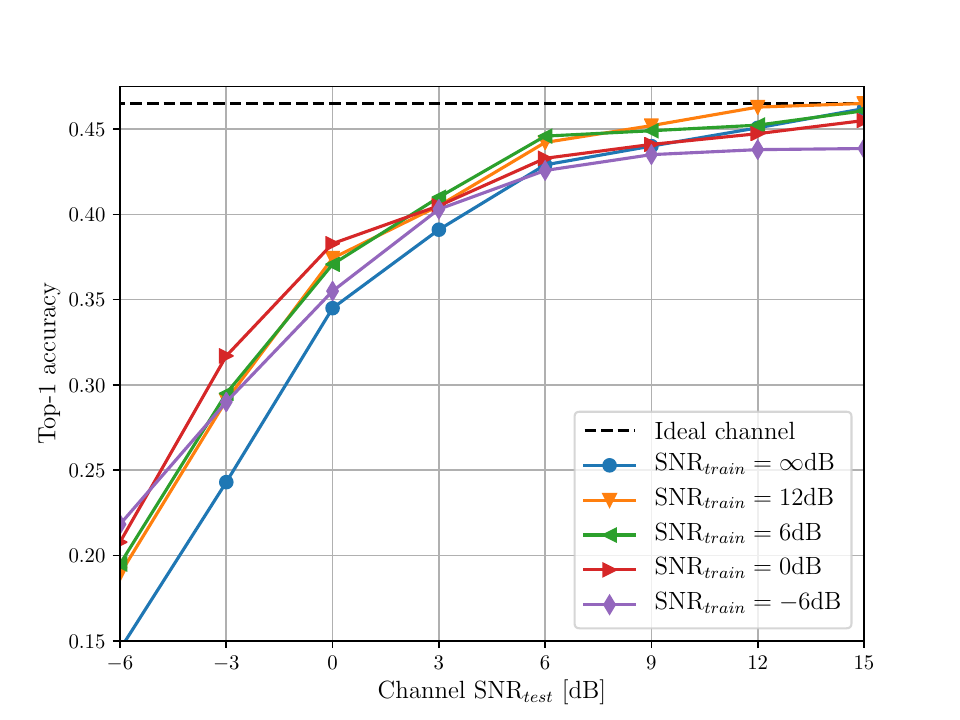}
    \caption{AWGN channel}
    \label{fig:results_graceful_degradation_cuhk_awgn}
    \end{subfigure}
    \begin{subfigure}[t]{0.49\textwidth}
        \centering
    \includegraphics[width=\textwidth]{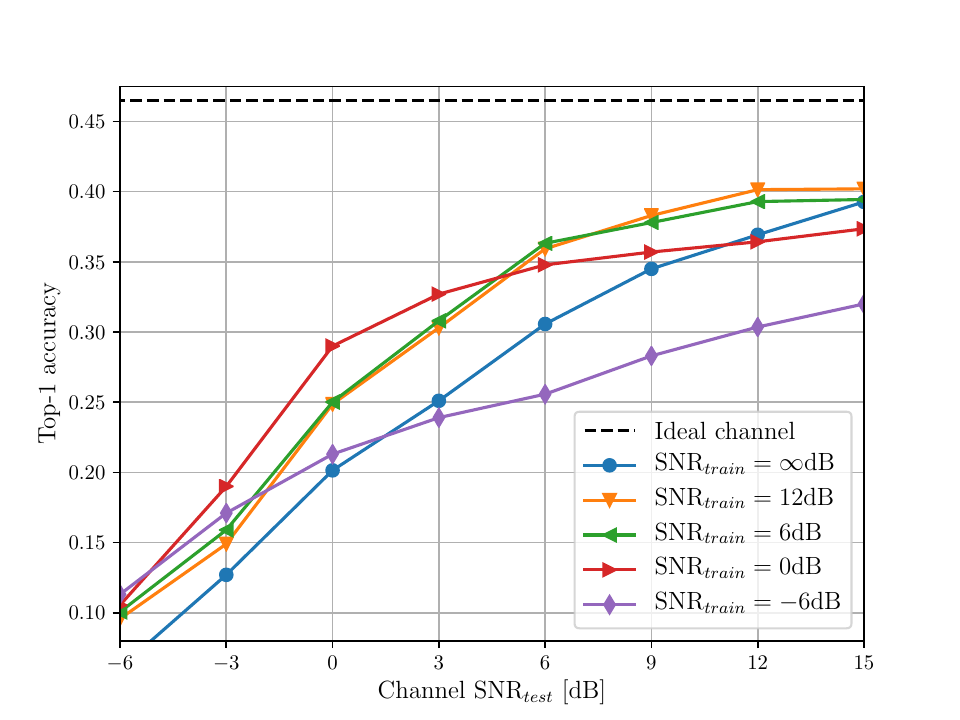}
    \caption{Fading channel}
    \label{fig:results_graceful_degradation_cuhk_fading}
    \end{subfigure}
    \caption{Accuracy achieved by the proposed JSCC AE scheme as a function of $\mathrm{SNR}_{test}$ for different $\mathrm{SNR}_{train}$ values for $B=64$. JSCC AE achieves graceful degradation with the channel SNR as opposed to the digital scheme, which suffers from the cliff effect. Models trained at moderate SNR$_{train}$ values achieve relatively good performance for a wide range of test SNRs values.}
    \label{fig:graceful_degradation}
\end{figure*}

In this section we evaluate the behaviour of our models on the CUHK03 dataset when the training and test SNRs do not match. In the experiments with the digital scheme, we assume that capacity-achieving channel codes are in use, and the quality of the channel is always estimated correctly. However, in practice, digital approaches suffer from the \textit{cliff effect}, which results in a sharp decrease in the performance when the channel condition is worse than the channel state, for which the channel code is designed. If the code rate is above the current channel capacity, it is known that true error probability converges to 1 \cite{gallager_comms}.

\begin{figure}[]
\begin{center}
\includegraphics[width=0.49\textwidth]{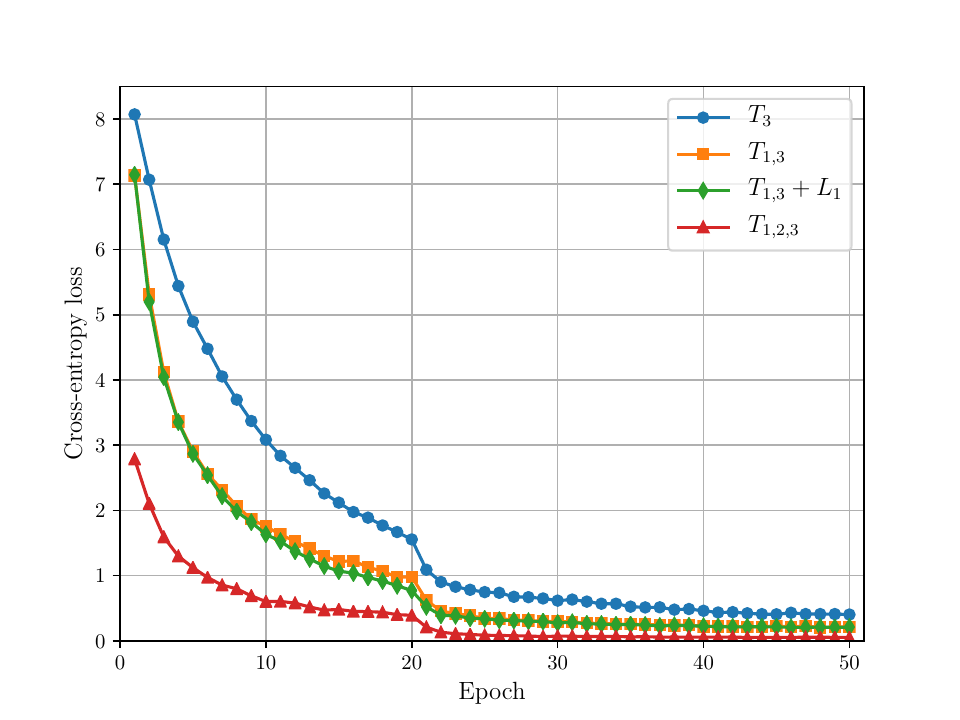}
\end{center}
   \caption{Comparison of training strategies through the evolution of the cross-entropy loss in the final joint training step. The proposed $T_{1, 2, 3}$ is superior to the alternative approaches.}
\label{fig:training_curves}
\end{figure}

On the other hand, unlike digital models, analog transmission schemes are known to achieve graceful degradation when we are interested in the end-to-end reconstruction quality \cite{jscc_image_dnn}; that is the average reconstruction quality smoothly decreases as the channel conditions become worse. This behaviour is quite beneficial, since we do not have to train multiple autoencoders, one for each channel SNR value, or even introduce channel estimation and feedback feature if the performance does not critically depend on applying the same training and testing SNRs. In the previous sections we showed the best possible accuracy for a specific SNR, which means each data point corresponds to a model trained specifically for that targer SNR. In Fig.~\ref{fig:graceful_degradation} we show that graceful degradation can be achieved with the proposed JSCC AE architecture, and it is not necessary to train a separate model for every SNR value. Instead, we can take a model trained with a moderate $\text{SNR}_{train}$ and apply it to a wide range of $\text{SNR}_{test}$ in the inference time at the expense of a moderate loss in accuracy. To the best of our knowledge, this is the first time graceful degradation is demonstrated for the inference as opposed to the average reconstruction quality that is typically considered in the literature.

 Note that the approach trained without noise ($\text{SNR}_{train}=\infty \mathrm{dB}$) is not robust against the channel noise. Therefore, its accuracy decreases much faster than for the networks trained under different noise levels, yet it still shows graceful degradation as the channel noise increases.

\subsection{Training Strategy}
\label{subs:training_strategy}

In this section we show the superiority of $T_{1, 2, 3}$ training strategy by comparing to the alternative training methods introduced in Section \ref{subs:approach1}. Note that, for the fairness of the comparison, we perform the first step of the training, which is the feature encoder pretraining, only once for $T_{1, 2, 3}$, $T_{1, 2}$ $T_{1, 3}$, and $T_{1, 3} + L_1$.

The evolution of the cross-entropy loss over training epochs of the final joint training phase for different training strategies is shown in Fig.~\ref{fig:training_curves}. In the experiment we used the bandwidth $B=64$ and SNR=$0\mathrm{dB}$.  The proposed three-step training allows to achieve much better final performance, as shown in Table \ref{tab:training_strategies}. Here, we also shown the top-5 recognition accuracy i.e., the correct match was listed within the top 5 ranklist elements, and the mean average precision (mAP), which are  standard evaluation measures for the retrieval tasks. Adding each training step increases the performance gradually, and there is a significant difference between $T_{3}$, and $T_{1, 3}$, as well as between $T_{1, 3}$ and $T_{1, 2, 3}$. Our three-step strategy outperforms all three alternatives by a large margin as it converges faster and achieves the smallest loss  after the last epoch. As expected, $T_{3}$ performs the worse, since it has to learn both the retrieval and robustness against the noise in a single training step with randomly initialized weights. Interestingly, the convergence of the $T_{1, 3} + L_1$ seems to slightly outperform the convergence of the $T_{1, 3}$, thanks to the additional loss term, which forces the reconstructed features to be similar to the original ones. However, while this seems to speed-up the convergence of the autoencoder network marginally, it does not affect the final performance. The reasonable performance of $T_{1,2}$ shows that $T_2$ allows the autoencoder to produce good reconstructions of the feature vectors under noisy environment, but the gap between $T_{1,2}$ and $T_{1,2,3}$ indicate the necessity of joint training phase, $T_3$, which maximizes the task performance. One may argue that our $T_{1, 2, 3}$ strategy is slower compared to the alternatives, nevertheless adding the autoencoder pretraining phase is negligible in comparison to the joint training phase ($\sim 3$min vs. $\sim 1$hr).
 
\begin{table}
{
\caption{\label{tab:training_strategies} Comparison of the retrieval performance for different training strategies.}
\centering
\resizebox{\columnwidth}{!}{
 \begin{tabular}{|c|c|c|c|}
 \hline
 Method & Top-1 accuracy & Top-5 accuracy & mAP\\ [0.5ex] 
 \hline\hline
$T_{3}$ & 0.225 & 0.409 & 0.195 \\
 \hline
$T_{1, 3}$ & 0.312 & 0.533 & 0.286 \\
 \hline
$T_{1, 3} + L_1$ & 0.317 & 0.536 & 0.287 \\
\hline
$T_{1, 2}$ & 0.330 & 0.557 & 0.306 \\
 \hline
$T_{1, 2, 3}$ & \textbf{0.392} & \textbf{0.602} & \textbf{0.351} \\

\hline
\end{tabular}
}
}
\end{table}

\begin{table*}
\caption{\label{tab:model_comparison} Person retrieval accuracy for the CUHK03 dataset achieved by different models at $\mathrm{SNR}=0\mathrm{dB}$ and $B=64$.}
\centering
 \begin{tabular}{|c|c|c|c|c|c|c|c|} 
 \hline
 Model & \# JSCC encoder layers & \# JSCC decoder layers & Activation & MSE & Top-1 accuracy & Top-5 accuracy & mAP \\ [0.5ex] 
 \hline\hline
A & 3 & 3 & Leaky ReLU & 0.204 & 0.382 & 0\textbf{.602} & 0.354 \\ 
 \hline
B & 3 & 2 & Leaky ReLU & 0.222 & 0.391 & 0.597 & 0.354 \\
 \hline
C & 3 & 4 & Leaky ReLU & 0.199 & 0.390 & 0.601 & 0.358 \\
 \hline
\textbf{D} & 2 & 3 & Leaky ReLU & 0.202 & \textbf{0.392} & \textbf{0.602} & \textbf{0.359} \\
 \hline
 D & 4 & 3 & PReLU & \textbf{0.181} & 0.383 & 0.589 & 0.343 \\
 \hline
E & 4 & 3 & Leaky ReLU & 0.208 & 0.383 & 0.598 & 0.356 \\
\hline
F & 1 & 1 & N/A & 0.207 & 0.387 & 0.592 & 0.352 \\
 \hline
 G & 2 & 2 & Leaky ReLU & 0.206 & 0.387 & 0.592 & 0.352 \\
 \hline
 H & 1 & 2 & Leaky ReLU & 0.207 & 0.386 & 0.593 & 0.353 \\
 \hline
 I & 2 & 1 & Leaky ReLU & 0.206 & 0.389 & 0.600 & 0.356 \\
 \hline
\end{tabular}
\end{table*}

\subsection{Comparison of Different Models}

In this section we present the results of architecture search for the JSCC autoencoder that resulted in the best performing model presented in Fig.~\ref{fig:autoencoder}. We considered 9 models designed as follows: both the JSCC encoder and the JSCC decoder are built of fully-connected layers, followed by the BN and activation layers. The only exceptions are the last layers in the JSCC encoder and the JSCC decoder which are without BN  and activations. The first layer of the JSCC encoder maps 2048-dimensional features to $2B$ real-valued symbols, which eventually forms a $B$ complex symbols transmitted over the channel. Similarly, the last layer of the JSCC decoder maps $2B$-dimensional vectors back to the original 2048-dimensional feature space and the remaining FC layers keep the dimension at $2B$. The evaluated architectures and results are shown in Table \ref{tab:model_comparison}. We select the models by starting from the baseline (denoted as A) from \cite{jankowski2019deep} and then removing or adding layers from the JSCC encoder and the JSCC decoder networks to explore the impact of depth on the overall performance.

For each model we trained the network according to the three-step strategy described in Section \ref{subs:approach1} and performed evaluation  on the CUHK03 dataset at $\text{SNR}=0\mathrm{dB}$, $B=64$.  We also show the mean squared error between the original feature vectors and their noisy reconstructions, after JSCC autoencoder pretraining $T_2$.
The results show that the differences between the models are marginal.  Model D, which corresponds to the architecture presented in Fig.~\ref{fig:autoencoder} and was used in the rest of the paper, performs slightly better than the others in terms of final retrieval performance. This model was selected also due to its low computational cost, as it  consists of only 5 fully-connected layers in total. We also used PReLU as the activation for the model variant D, and observed that even though it achieves better MSE in step $T_2$, it fails to provide a good generalization capabilities in the final step, as it overfits to the data. Please note that the model F, does not have the activation function, as the only layers in both the encoder, and the decoder are the last layers, therefore, as described above, the activation and BN are removed.

\section{Conclusions}

In this work, we have introduced the image retrieval problem over wireless channels in the context of the edge network, where wireless edge devices send queries of images over a bandwidth and power limited channel to an edge server that stores the image database, also called the gallery. We first introduced a digital approach, which is based on a novel retrieval-oriented deep image compression scheme, and applied it to feature vectors obtained from the feature encoder. Next, we proposed a JSCC-based scheme, where feature vectors are directly mapped to the channel symbols and decoded at the receiver. We showed the latter approach not only achieves a superior retrieval accuracy at a target channel SNR, but also provides graceful degradation with the test SNR when it does not match the training SNR. We further introduced JSCC FC, which is a simplified version of the proposed model and showed that decoding is necessary at the receiver to mitigate the effects of channel impairments. We also proposed a novel strategy for training our JSCC scheme, that can be adapted to other machine learning applications performed over noisy channels. Our strategy  achieves superior performance for training the JSCC scheme. We have also performed an extensive ablation study of different architectures and training strategies and compared the alternatives under various performance measures for a wide range of different channel conditions. The results show the superiority of the proposed architecture and the joint training approach.

\bibliographystyle{IEEEbib}
\bibliography{refs}

\ifCLASSOPTIONcaptionsoff
  \newpage
\fi

\begin{IEEEbiography}[{\includegraphics[width=1in,height=1.25in,clip,keepaspectratio]{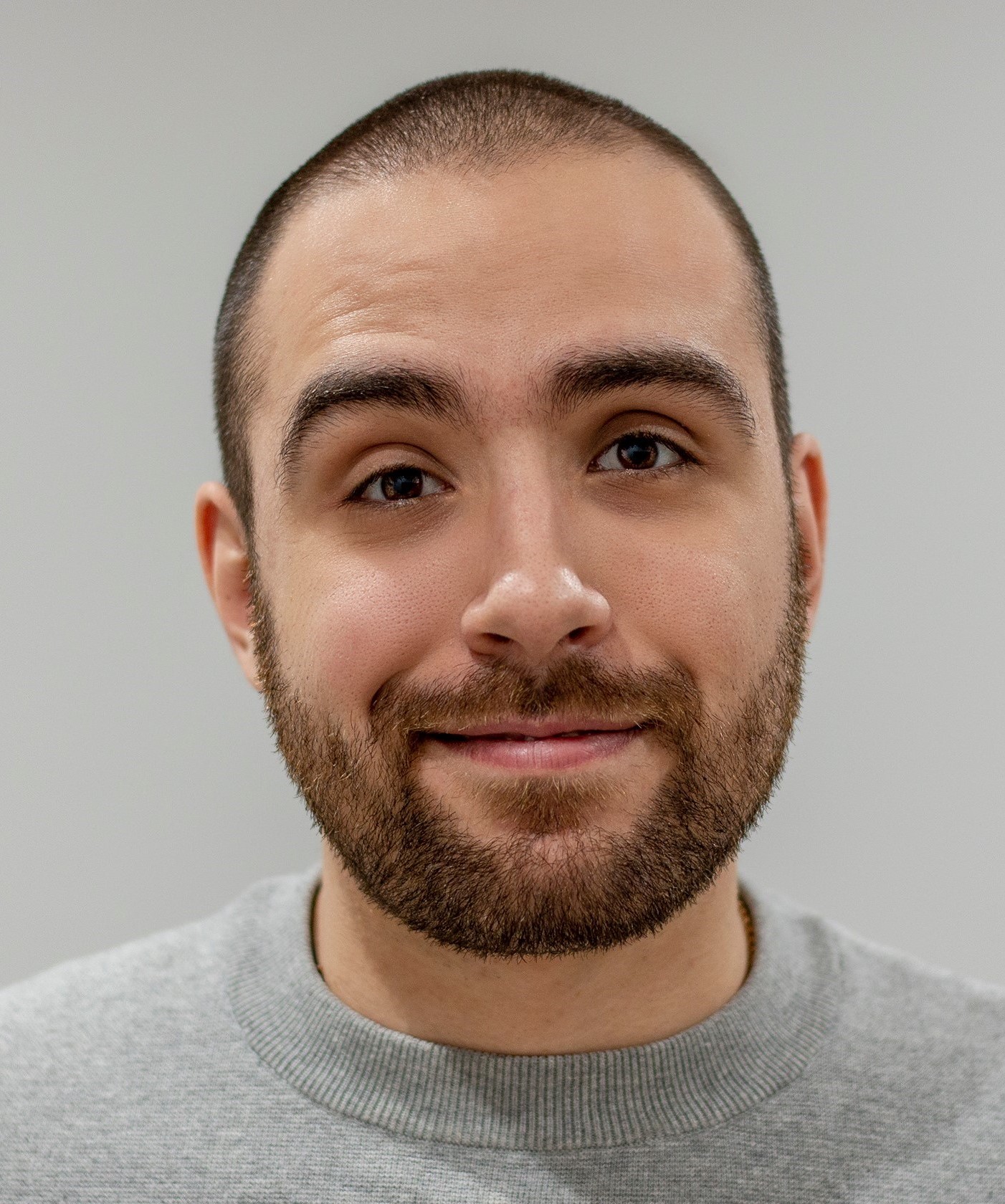}}]{Mikolaj Jankowski}
[S’18] received his BSc degree in Control and Robotics from Warsaw University of Technology in 2016 and MSc in Control Systems from Imperial College London in 2018. Currently he is a PhD student at Intelligent Systems and Networks Group, under the supervision of Prof. G\"und\"uz and Dr. Mikolajczyk. His main research areas are machine learning, computer vision, image processing and information theory. He is currently working on data compression optimized for various machine learning tasks and applying machine learning to improve wireless communications.
\end{IEEEbiography}

\begin{IEEEbiography}[{\includegraphics[width=1in,height=1.25in,clip,keepaspectratio]{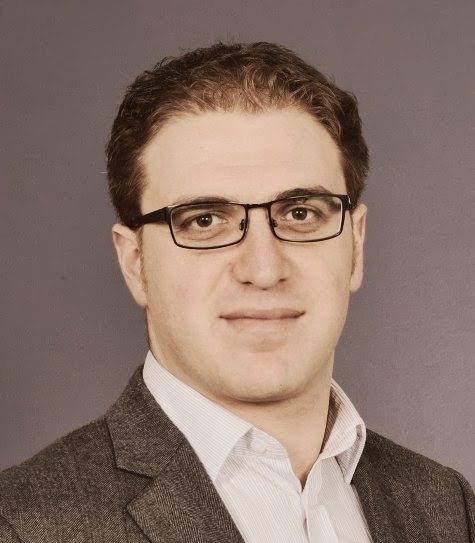}}]{Deniz G\"und\"uz}
[S’03-M’08-SM’13] received his Ph.D. degree from NYU Tandon School of Engineering (formerly Polytechnic University). Currently, he is a Professor of Information Processing at Imperial College London, UK, and serves as the deputy head of the Intelligent Systems and Networks Group. He is also a part-time faculty member at the University of Modena and Reggio Emilia. He served in various editorial roles for the IEEE Transactions on Communications, IEEE Journal on Selected Areas in Communications, IEEE Transactions on Green Communications and Networking, and IEEE Transactions on Wireless Communications. He is a Distinguished Lecturer for the IEEE Information Theory Society (2020-21). He is the recipient of the IEEE Communications Society - Communication Theory Technical Committee (CTTC) Early Achievement Award in 2017 and a Starting Grant of the European Research Council (ERC) in 2016. His research interests lie in the areas of communications, information theory, machine learning, and privacy. 
\end{IEEEbiography}


\begin{IEEEbiography}[{\includegraphics[width=1in,height=1.25in,clip,keepaspectratio]{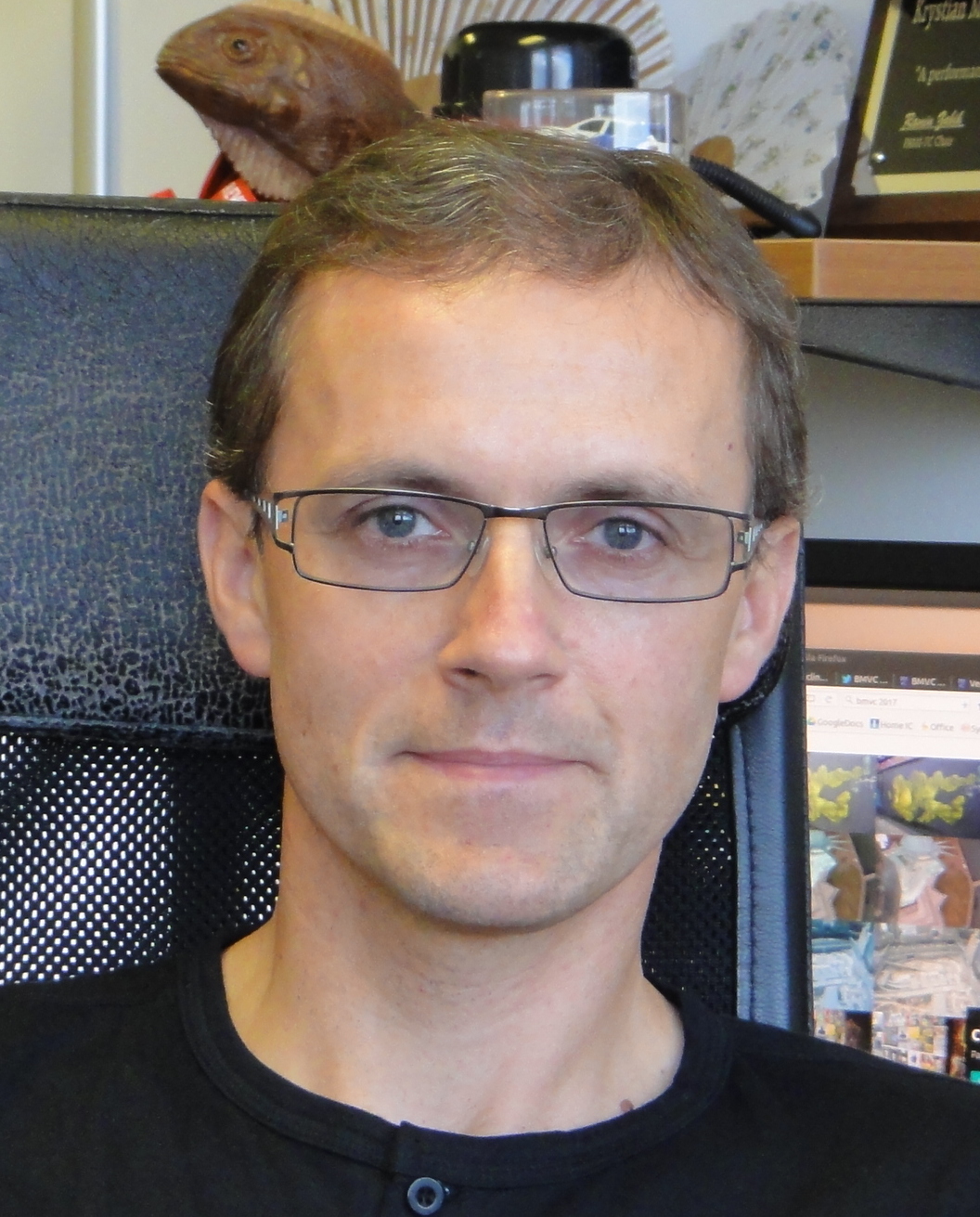}}]{Krystian Mikolajczyk}
[S’99-M’02-SM’09]  is an Associate Professor at Imperial College London. He completed his PhD degree at the Institute National Polytechnique de Grenoble and held a number of research positions at INRIA, University of Oxford and Technical University of Darmstadt, as well as faculty positions at the University of Surrey, and Imperial College London. His main area of expertise is in image and video recognition, in particular methods for image representation and learning. He has served in various roles at major international conferences co-chairing British Machine Vision Conference 2012, 2017 and IEEE International Conference on Advanced Video and Signal-Based Surveillance 2013. In 2014 he received Longuet-Higgins Prize awarded by the Technical Committee on Pattern Analysis and Machine Intelligence of the IEEE Computer Society.
\end{IEEEbiography}




\end{document}